# Magneto-optical Skyrmion for manipulation of arbitrary light polarization

**Authors:** *Xinghong Chen[1†], Xing-Xiang Wang[2†], Guanjie Zhang[1], Xiao Hu[2,3], Wei Tao[1*], Tomohiro Amemiya[4*], Yifei Mao[1*]*

**Affiliations:**

[1]State Key Laboratory of Submarine Geoscience, School of Automation and Intelligent Sensing, Shanghai Jiao Tong University, Shanghai 200240, China.

[2]Institute for Quantum Science and Technology, Shanghai University, Shanghai 200444, China

[3]Department of Physics, Shanghai University, Shanghai 200444, China

[4]School of Engineering, Institute of Science Tokyo, Tokyo 152-8552, Japan.

[*]Correspondence to:

maoyifei@sjtu.edu.cn; amemiya.t.ab@m.titech.ac.jp; taowei@sjtu.edu.cn

[†]These authors contributed equally to this work.

## Abstract

Dynamic manipulation of arbitrary light polarization is of fundamental importance for versatile optical functionalities, yet realizing such full-Poincaré-sphere control within compact nanophotonic architectures remains a formidable challenge. Here, we theoretically propose and numerically demonstrate a magneto-optical (MO) skyrmion platform enabling full polarization control of cavity eigenmodes. We reveal the correspondence between the near-field wavefunctions of degenerate dipoles and far-field polarization. By applying multidirectional magnetic fields to MO photonic crystals, we achieve any complex superposition of orthogonal eigenmodes, thereby realizing arbitrary far-field polarization. This mapping manifests as a skyrmion with a topological charge of 2, guaranteeing coverage of the entire Poincaré sphere. Our theoretical model shows excellent agreement with finite-element simulations. Furthermore, we realize bound states in the continuum (BICs) with dynamically tunable polarization textures and demonstrate high-performance polarization-selective emission and transmission. This work establishes a topological paradigm for precise polarization shaping, offering new avenues for advanced optical communication and sensing.

**Introduction**

Polarization of light is of fundamental importance, and the ability to control light polarization finds applications in numerous fields such as sensing[1-3], imaging[4-8], and optical and quantum communications[9-13]. Conventionally, a laser beam can be converted into desired polarizations after passing through wave plates exhibiting the birefringence effect[14-18]. While this technique is commonly used in almost every free-space optics platform nowadays, direct manipulation of the inherent polarization of the light source is desired for modern optical applications. Recent studies show that polarized emissions can be obtained from the eigenstates of elaborately designed photonic structures[19-26]. One study has proposed that any static polarization can be achieved by adjusting the spacing between two one-dimensional photonic crystals[19]. Dynamic modulations of the photonic structures are also achieved through extra efforts. For instance, the crystal orientation of liquid crystal materials can be tuned electrically, which modifies the refractive index distribution and consequently changes the eigenmodes of the photonic systems composed of liquid crystals[20]. So far, such approaches often involve bulky structures, limiting their potential for integration. Magneto-optical (MO) materials hosting Faraday's effect are promising candidates for achieving dynamic control of polarized emission[27,28]. In fact, switching between linear and circular polarization has been demonstrated recently[29]. However, achieving dynamic tuning of arbitrary polarization remains challenging.

To realize the full polarization control, one needs a surjective mapping from a continuous controllable parameter space to the Stokes space (or Poincaré sphere), which is reminiscent of a skyrmion[30-36]. In this work, we conceptualize and demonstrate a MO skyrmion platform that enables arbitrary polarization control of cavity eigenmodes, where the skyrmion is defined by the mapping from the direction of the magnetic field to the Stokes space of the far-field polarization. The skyrmion number corresponds to the number of times that generic polarization states are achieved, whereas a finite number of critical states may be attained with different non-zero multiplicity. In words, the realization of the full polarization control can be converted to the realization of a non-trivial skyrmion. By constructing photonic crystal (PhC) slabs with MO materials and subjecting them to an external magnetic field, we remove the degeneracy of a pair of orthogonal dipole modes. We show that the way of superposition of the dipole modes is bound to the orientation of the magnetic field and determines the far-field polarization on the Poincaré sphere. As the magnetic field rotates and traverses all possible directions in the real space, the polarization state wraps twice on the Poincaré sphere, forming a skyrmion with a topological charge of 2, as shown in FIG. 1. We have derived theoretically the relationship between the far-field polarization of the eigenmodes and the direction of the applied magnetic field, which shows excellent agreement with the simulation results. Furthermore, we realize bound states in the continuum (BICs) by tuning the magnetic field strength. Crucially, the polarization textures of BICs are dynamically tunable, holding promise for dynamically controllable polarized lasers. Specifically, under a purely out-of-plane magnetic field, the polarization near the singularity of the far field becomes elliptical rather than linear, whereas under a purely in-plane magnetic field, the overall polarization texture near the Γ point is stretched along the magnetic field direction. Additionally, we demonstrate polarization-selective emission and highly efficient dichroic transmission characterized by significant linear dichroism (LD > 0.8) and circular dichroism (CD ≈ 1). Our work provides a novel approach for precise and efficient polarization control, offering promising prospects for advancements in optical communications, sensing, and imaging technologies.

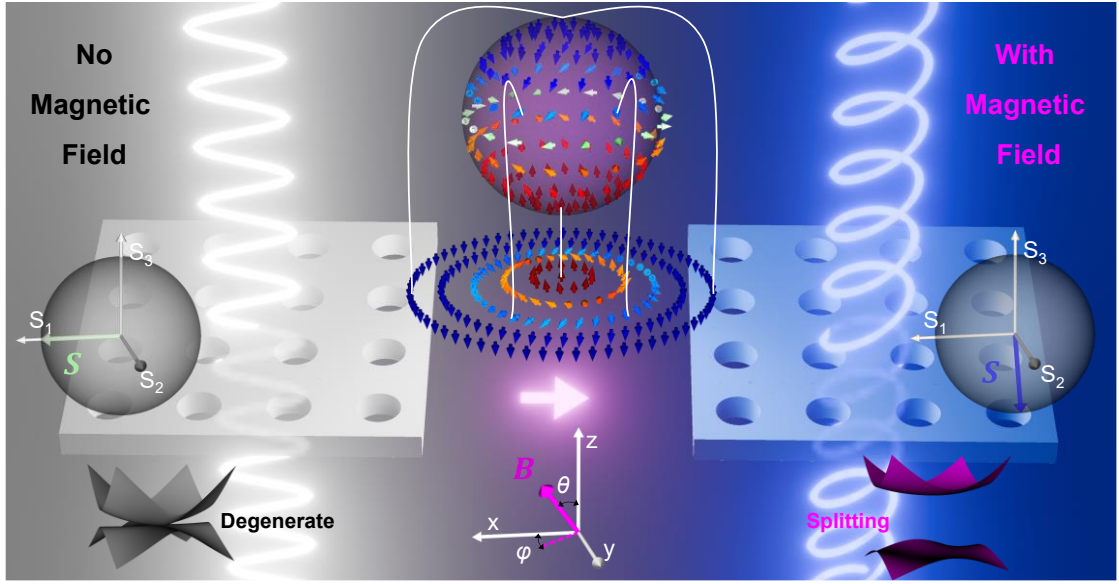

**FIG. 1.** Illustrations of polarization control via the direction of magnetic field. Without an external magnetic field, the dipolar eigenmodes of the PhC slab are linearly polarized and exhibit band degeneracy. Applying a magnetic field lifts the degeneracy and induces elliptically polarized eigenmodes when both in-plane and out-of-plane components are present. As the magnetic field direction completes a full rotation in real space, the polarization state winds twice on the Poincaré sphere, forming a skyrmion with a topological charge of 2.

**Results:**

In the present work, we investigate a MO PhC slab as shown in FIG. 2(a). The lattice constant of the PhC slab is $a = 500$ nm, the radius of the air holes is $r = 150$ nm, and the thickness of the slab is $t = 600$ nm. In a MO material, the relative permittivity tensor $\epsilon$ is

$$\epsilon(\delta) = \begin{bmatrix} \epsilon_{zf} & i\delta \cos\theta & -i\delta \sin\theta \sin\varphi \\ -i\delta \cos\theta & \epsilon_{zf} & i\delta \sin\theta \cos\varphi \\ i\delta \sin\theta \sin\varphi & -i\delta \sin\theta \cos\varphi & \epsilon_{zf} \end{bmatrix} \quad (1)$$

where $\varphi$ and $\theta$ are the azimuthal angles of the external magnetic field in the spherical coordinate shown in FIG. 1, and $\delta$ is a small quantity related to the intensity of the magnetic field ($\delta = 0$ when magnetic field is zero)[37]. In the present work, $\epsilon_{zf}$ is set to be 4. This PhC holds $C_{4v}$ symmetry when there is no external magnetic field. In this context, as shown in the band structure illustrated in FIG. 2(b), there is a pair of degenerate dipole states $|p_x\rangle$ and $|p_y\rangle$ at $\Gamma$ point of the Brillouin zone. This degeneracy is guaranteed by the 2D irreducible representation of $C_{4v}$ point group. When the PhC slab is subjected to an external magnetic field, the time reversal symmetry is no longer present, and the $C_{4v}$ symmetry is either reduced to $C_4$ (when the magnetic field is perpendicular to the PhC slab) or completely broken. In both cases, the 2D representation is not irreducible anymore, which leads to the frequency splitting of the degenerate modes. FIG. 2(c) shows the case that the two-fold degeneracy is gapped by a vertical magnetic field.

Mathematically, the frequency splitting corresponds to an orthogonal direct sum decomposition of the space spanned by $\{|p_x\rangle, |p_y\rangle\}$. The orthonormal basis of this decomposition, namely, the split pair of eigenstates can be written as

$$|e_1\rangle = c_x|p_x\rangle + c_y|p_y\rangle$$

$$|e_2\rangle = c_y^*|p_x\rangle - c_x^*|p_y\rangle, \qquad (2)$$

with the orthonormal conditions $\langle p_\alpha|p_\beta\rangle_{\text{inner}} \equiv \langle p_\alpha|\epsilon|p_\beta\rangle = \int_{\text{u.c.}} d\mathbf{r} \mathbf{E}_{p_\alpha}^* \cdot \epsilon \mathbf{E}_{p_\beta} = \delta_{\alpha\beta}$ and $\langle e_m|e_n\rangle_{\text{inner}} = \delta_{mn}$ ($\delta_{\alpha\beta}$ and $\delta_{mn}$ are the Kronecker's deltas). We note that, the far-field radiation behavior is deeply related to the near-field wavefunctions of the eigenstates. Specifically, in the present PhC structure with $C_4$ rotational symmetry, we prove that for the TM dipole modes, the far field at the $\Gamma$ point obeys a one-to-one correspondence that the electric field components follow $E_{f,|e_1\rangle} \propto c_x\hat{x} + c_y\hat{y}$, where the coefficients of $x$ and $y$ linear polarized components are exactly same as the combination coefficients of the dipole eigenstates (see Supplemental Material S1 for detailed derivation[38]). Namely, a near-field dipole/vortex gives a specific far-field linear/circular state. This indicates that the polarization of the far-field states can be arbitrarily manipulated as long as one can control the combination coefficients $c_x$ and $c_y$ of the near-field dipole states, which, from an intuitive perspective, are determined by the magnetic field.

In what follows, we show that any combination of the near-field dipoles, and thus all far-field polarizations, can be achieved twice by sweeping the magnetic field with a fixed intensity over all possible orientations in the real space, and thus all far-field polarizations can be achieved twice by sweeping the magnetic field with a fixed intensity over all possible orientations in the real space (except for the pure circular polarizations at the poles of the Poincaré sphere, which are achieved only once). In other words, the mapping from the magnetic field direction to the Poincaré sphere ($S^2 \to S^2$) is a two-sheeted ramified cover, which defines a skyrmion structure with a skyrmion number of 2. When the magnetic field is weak, this can be demonstrated by using a degenerate perturbative approach.

The master equation derived from Maxwell's equations reads

$$\mathcal{H}(\delta)\mathbf{E} = \epsilon^{-1}(\delta)\nabla \times \nabla \times \mathbf{E} = \frac{\omega^2}{c^2}\mathbf{E} \tag{3}$$

with $\mathcal{H} = \epsilon^{-1}\nabla \times \nabla \times$ the Hamiltonian.

The Hamiltonian can be divided into two parts $\mathcal{H} = \mathcal{H}_0 + \mathcal{H}'$, where $\mathcal{H}_0 = \epsilon^{-1}(0)\nabla \times \nabla \times$ and $\mathcal{H}'$ is the perturbation part contributed by the external magnetic field. Here, we use $\{|p_x\rangle, |p_y\rangle\}$ as the unperturbed basis.

To be noticed that $\nabla \times \nabla \times \mathbf{E}_{p_\alpha} = \epsilon(0)(\omega_0^2/c^2)\mathbf{E}_{p_\alpha}$ holds for both $\alpha = x$ and $\alpha = y$, the elements of the $2 \times 2$ perturbation Hamiltonian are thus

$$\begin{aligned}\mathcal{H}'_{\alpha\beta} &= \langle p_\alpha|\epsilon(0)[\mathcal{H}(\delta) - \mathcal{H}(0)]|p_\beta\rangle \\ &= \langle p_\alpha|\epsilon(0)[\epsilon^{-1}(\delta) - \epsilon^{-1}(0)]\nabla \times \nabla \times |p_\beta\rangle \\ &= \frac{\omega_0^2}{c^2}\langle p_\alpha|\eta(\delta)|p_\beta\rangle \\ &= \int_{\text{u.c.}} \mathbf{E}^*_{p_\alpha} \cdot \eta(\delta)\mathbf{E}_{p_\beta}d\mathbf{r},\end{aligned} \tag{4}$$

where $\eta(\delta) = \epsilon(0)[\epsilon^{-1}(\delta) - \epsilon^{-1}(0)]\epsilon(0)$. Here, the dipole modes which we focus on are TM modes. By considering the symmetries of the electric components of these modes (see Supplemental Material S2[38]), the integral in Eq. (4) can be simplified. Defining $A_i^\alpha = (\omega_0^2/c^2)\int_{\text{slab}} d\mathbf{r}|E_{p_\alpha,i}|^2$ and $B_\pm = (\omega_0^2/c^2)\int_{\text{slab}} d\mathbf{r}\left(E^*_{p_x,x}E_{p_y,y} \pm E^*_{p_x,y}E_{p_y,x}\right)$, whose values are numerically given by the simulated unperturbed states, the Hamiltonian can be written as

$$\mathcal{H}' = \frac{\delta\epsilon_{\text{zf}}}{\epsilon_{\text{zf}}^2 - \delta^2}\begin{bmatrix} h_x & h_{xy} \\ h^*_{xy} & h_y \end{bmatrix}, \tag{5}$$

with $h_\alpha = \delta(A_x^\alpha + A_y^\alpha) + \delta \sin^2\theta (A_z^\alpha - A_x^\alpha \cos^2\varphi - A_y^\alpha \sin^2\varphi)$, and $h_{xy} = -\delta B_+ \cos\varphi \sin\varphi \sin^2\theta - i\epsilon_{zf} B_- \cos\theta$. The frequency splitting under the magnetic field can be obtained by diagonalizing the perturbation Hamiltonian Eq. (5), and the components of the eigenvectors of $\mathcal{H}'$ gives the combination coefficients $c_x$ and $c_y$ in Eq. (1).

In order to investigate the relation between the magnetic field orientation $(\varphi, \theta)$ and the combination coefficients $c_x$ and $c_y$ of the dipole modes in the split eigenstate, we calculate the normalized Stokes parameters $\mathbf{S}(\varphi, \theta) = (S_1, S_2, S_3)$ of $|e_1\rangle$ with $|\mathbf{S}| = 1$ from both the first principle simulation and the theoretically derived Hamiltonian (8). The components of the $\mathbf{S}$ are defined by

$$S_1 = |c_x|^2 - |c_y|^2$$
$$S_2 = 2\mathrm{Re}(c_x c_y^*)$$
$$S_3 = -2\mathrm{Im}(c_x c_y^*) \tag{6}$$

where Re and Im denote taking the real and imaginary parts of the complex number, respectively. As shown in FIG. 2(d) for the first principle simulation results, it is observed that by varying the magnetic field direction on the $+x$ hemisphere, any possible set of $S_1$, $S_2$, and $S_3$, namely, any far-field polarization states can be achieved exactly once. In parallel, the results obtained via the theoretical model are shown in FIG. 2(e), which are in fully consistency with the simulation. Due to the mirror symmetry with respect to the $yz$-plane, same results can be obtained when the magnetic field orients the $-x$ hemisphere. Therefore, the function $\mathbf{S}(\varphi, \theta)$ is a two-sheeted covering map corresponding to a skyrmion structure as shown in the upper part of FIG. 2(f), where the sphere stands for the direction of the magnetic field $\mathbf{B}/|\mathbf{B}|$ and the arrows on the sphere represent the Stokes parameters. The skyrmion number $n_s$ can be calculated by using

$$n_s = \frac{1}{4\pi} \int \mathbf{S} \cdot \left(\frac{\partial \mathbf{S}}{\partial x} \times \frac{\partial \mathbf{S}}{\partial y}\right) dxdy = 2, \tag{7}$$

where the Cartesian coordinate $(x, y) = [\theta \cos(\pi - \varphi), \theta \sin(\pi - \varphi)]$ is obtained by flattening the sphere in the way shown in FIG. 2(f).

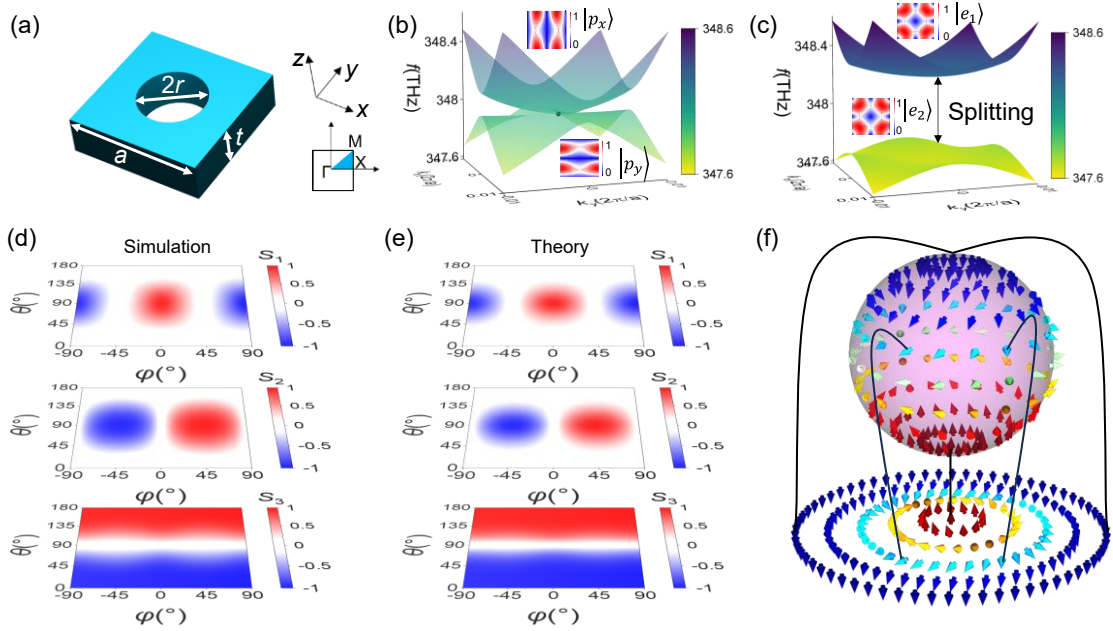

**FIG. 2.** (a) Schematic of a MO PhC slab. Here, $a$ is the lattice constant, $r$ the air-hole radius, and $t$ the slab thickness. (b, c) Band structures of two modes without and with the magnetic field ($\theta = 0°$), respectively. The insets show the corresponding electric-field intensity distributions. (d, e) Simulated and theoretical Stokes parameters of the far-field polarization of mode $|e_1\rangle$ at the Γ point under different magnetic field orientations. At $\theta = 0°$ and $180°$, the far-field polarization is independent of $\varphi$. (f) MO polarization skyrmion reconstructed from simulation results (d).

Next, we simulated the far-field polarization at Γ point of $|e_1\rangle$ under different magnetic field orientations with a fixed $\delta = 0.08$, as shown in FIG. 3(a). When an in-plane magnetic field is applied, the resulting polarization is linear, with its orientation varying according to the magnetic field direction. In this context, $|e_1\rangle$ is merely a real-valued superposition of the $|p_x\rangle$ and $|p_y\rangle$. Under a purely out-of-plane magnetic field, circular polarization is obtained. When both in-plane and out-of-plane components are present in the magnetic field, the in-plane component determines the polarization angle, whereas the out-of-plane component governs the ellipticity—the larger the out-of-plane field is applied, the lower the ellipticity. The simulated results of $|e_2\rangle$ under different magnetic field directions are orthogonal to $|e_1\rangle$, as shown in the Supplemental Material S3[38]. Additionally, in this structure, other degenerate dipole modes were also investigated. Their far-field polarizations under different magnetic fields were simulated (Supplemental Material S4[38]), and full polarization control can also be achieved. This indicates that our design enables arbitrary polarization control across multiple wavelengths. Furthermore, we calculated the far-field polarization under different magnetic field orientations based on our theoretical model, as shown in FIG. 3(b), which exhibits excellent agreement with the simulation results. We further examined the relationship between the in-plane magnetic field direction and the polarization angle, as shown in FIG. 3(c), where the simulation data are found to closely follow the theoretical curve.

When the magnetic field is oriented along the in-plane x-direction, the far-field radiation at the Γ point exhibits x-linear polarization as shown in FIG. 3(d). In addition to affecting the polarization at the Γ point, the applied magnetic field also influences the polarization in its vicinity. When only an out-of-plane magnetic field is applied in the $+z$ direction, the polarization near the Γ point becomes left-handed circular polarization (LCP), while farther away from the Γ point, the influence of the wavevector results in elliptical polarization, with the major axis orientation determined by the wavevector, as shown in FIG. 3(e). This further indicates that the out-of-plane magnetic field primarily affects the ellipticity of the polarization. Moreover, when equal in-plane and out-of-plane magnetic fields are applied ($\theta = 45°$, $\varphi = 45°$), the polarization near the Γ point is left-handed elliptical with its major axis oriented at 45°, and as the distance from the Γ point increases, both the major axis direction and the ellipticity are again modified by the wavevector. As the wavevector moves away from the Γ point, both the orientation of the major axis and the ellipticity are altered due to the influence of the wavevector, as shown in FIG. 3(f). In addition, the far-field polarization at different magnetic field strengths is discussed in Supplemental Material S5[38].

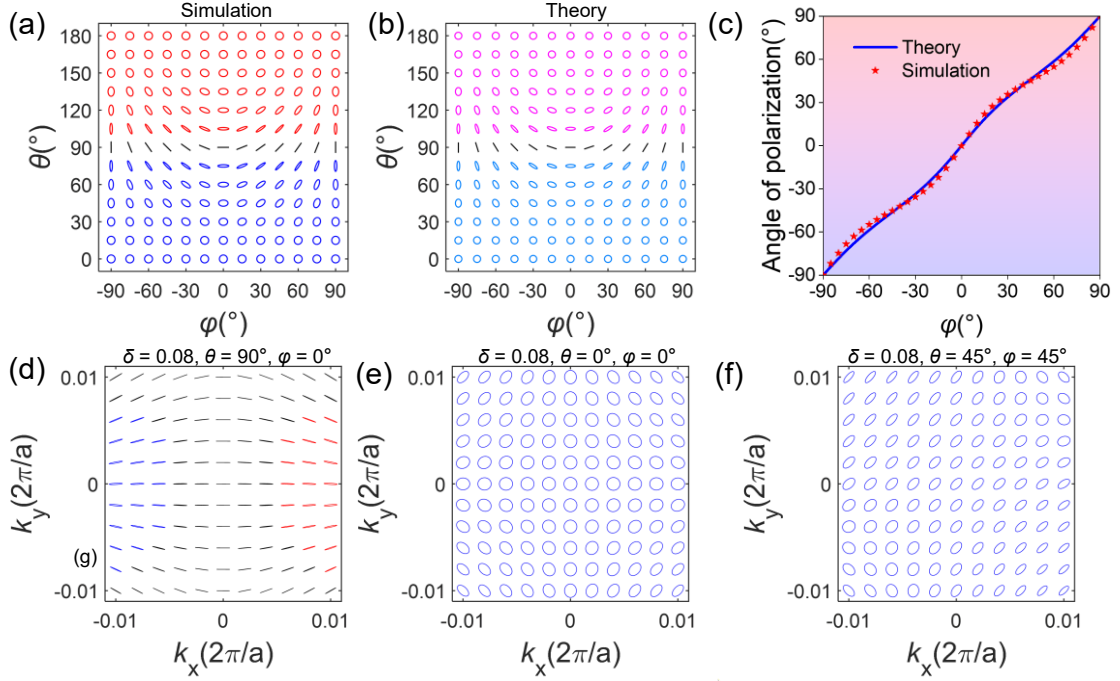

**FIG. 3.** Far-field polarization characteristics of mode $|e_1\rangle$ at the $\Gamma$ point under different magnetic field conditions. (a, b) Simulated and theoretical polarization ellipses of the far field for $|e_1\rangle$ at the $\Gamma$ point under different magnetic field orientations. Red and magenta represent right-handed polarization, blue and cyan represent left-handed polarization, and black represents linear polarization. (c) Simulated and theoretical results of the polarization angle when only an in-plane magnetic field is applied. (d–f) Far-field polarization distributions for $\delta = 0.08$ at different magnetic field orientations: (d) $\theta = 90°$, $\varphi = 0°$; (e) $\theta = 0°$ and $\varphi$ is arbitrary at $\theta = 0°$ (f) $\theta = 45°$, $\varphi = 45°$.

We investigate the evolution of the complex frequencies and topological characteristics of $|e_{1,2}\rangle$ under varying magnetic field configurations. The magnetic field lifts the degeneracy and modifies the $Q$ factors, leading to the formation of BICs with distinct polarization textures. First, under an out-of-plane magnetic field ($\theta = 0°$), the time-reversal symmetry breaking lifts the degeneracy at the $\Gamma$ point, causing a frequency splitting proportional to the field strength $\delta$, as shown in FIG. 4(a). Crucially, the imaginary part of the eigenfrequency for $|e_2\rangle$ vanishes at $\delta = 0.32$, indicating a divergence of the $Q$ factor and the formation of a BIC with a topological charge of $-1$[29]. Unlike conventional symmetry protected BICs with linear polarization, this magnetically induced BIC exhibits a polarization singularity surrounded by elliptical polarization[39].

In the case of an in-plane magnetic field ($\theta = 90°$), we observed that the far-field polarization textures of BICs are dependent on the orientation angle $\varphi$, as shown in FIG. 4(b)-(c). For $\varphi = 0°$, the $Q$ factors of $|e_{1,2}\rangle$ diverge along distinct trajectories, forming BICs at $\delta = 0.83$ (topological charge of $+1$) and $\delta = 0.95$ (topological charge of $-1$), respectively. Notably, the overall polarization texture near the $\Gamma$ point of $|e_1\rangle$ is stretched along $k_x$. We also simulated the far-field polarization of $|e_2\rangle$ at $\delta = 0.95$, as presented in the Supplemental Material S5[38]. Changing the field orientation to $\varphi = 45°$ shifts the BIC condition for $|e_1\rangle$ to $\delta = 0.49$, elongating the overall polarization texture near the $\Gamma$ point along $k_x = k_y$. Comprehensive simulations for various in-plane orientations reveal a general rule: the overall polarization texture near the $\Gamma$ point is consistently stretched along the direction of the applied magnetic field (see Supplemental Material S5[38]). This demonstrates a novel capability to dynamically manipulate the far-field polarization texture of BICs via the magnetic field feature not previously achieved. Conversely, under a mixed magnetic field configuration ($\theta = 45°$, $\varphi = 45°$), the total breaking of symmetries (in-plane and out-of-plane symmetries) prevents the formation of infinite-$Q$ modes (see Supplemental Material S6[38]).

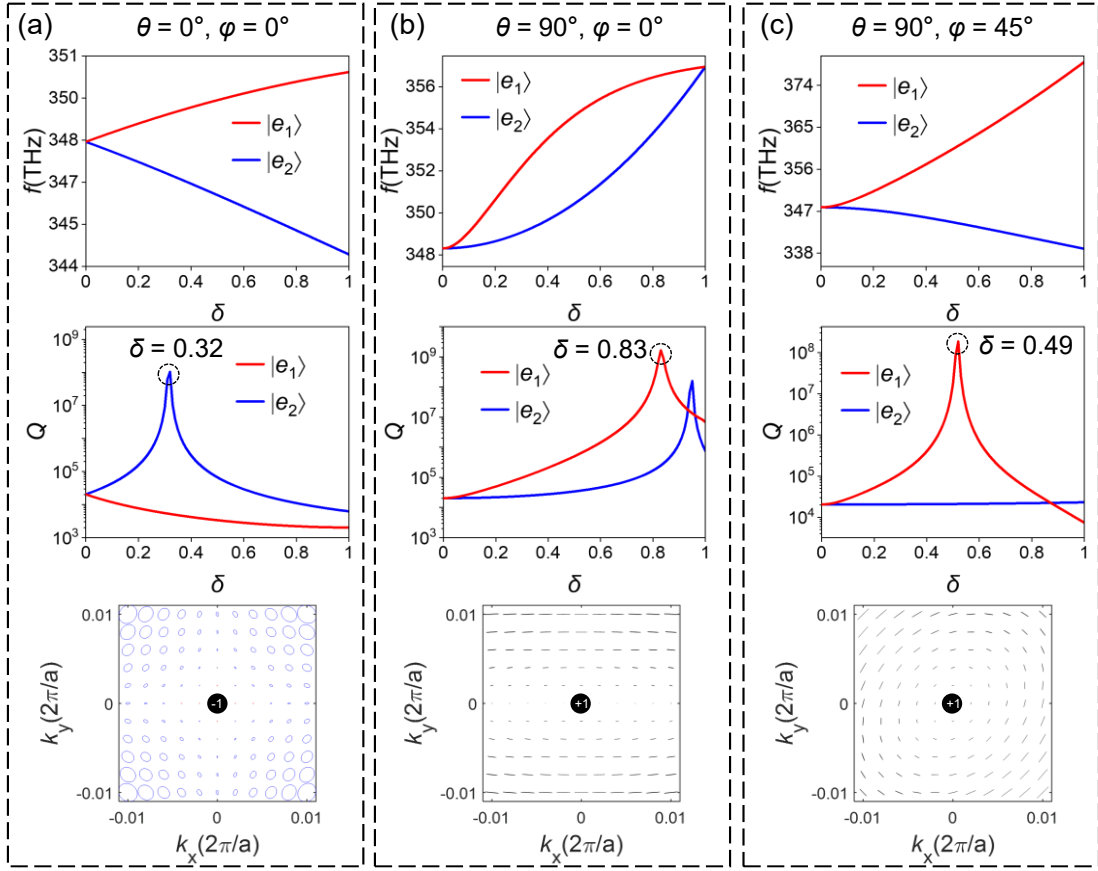

**FIG. 4.** Evolution of complex frequencies and far-field polarization textures under varying magnetic fields. The top, middle, and bottom panels in each column display the real frequency, $Q$ factor, and far-field polarization of the BIC, respectively. (a) Out-of-plane field ($\theta = 0°$). The polarization corresponds to $|e_2\rangle$ at $\delta = 0.32$. (b) In-plane field along the x-axis ($\theta = 90°$, $\varphi = 0°$). Polarization of $|e_1\rangle$ at $\delta = 0.83$. (c) In-plane field along 45° ($\theta = 90°$, $\varphi = 45°$). Polarization of $|e_1\rangle$ at $\delta = 0.49$.

To illustrate the potential applications of our MO PhCs in polarization control, we analyzed the emission and transmission spectra to evaluate the CD and LD[40-42]. Under an out-of-plane magnetic field ($\theta = 0°$), the device exhibits strong chiral characteristics, as shown in FIG. 5(a). The emission spectrum shows distinct peaks for LCP and right-handed circular polarization (RCP) light at 348.21 THz and 347.66 THz, respectively, suggesting potential as a circularly polarized laser. Correspondingly, the transmission analysis yields a giant $CD \approx -0.99$ at 348.22 THz. Here, CD is defined as $CD = \frac{T_L - T_R}{T_L + T_R}$, where $T_{L(R)}$ is the normalized transmission spectra under LCP (RCP) illumination. This value approaches the theoretical limit of $-1$, indicating high efficiency for chiral sensing applications. Under in-plane magnetic fields ($\theta = 90°$), the device demonstrates strong linear dichroism. For a field along the x-axis ($\varphi = 0°$), we observe splitting between x- and y-polarized emissions. A high $LD_{0°}$ is achieved at 348.01 THz, with a value of 0.84. $LD_{0°}$ is defined as $LD_{0°} = \frac{T_{0°} - T_{90°}}{T_{0°} + T_{90°}}$, where $T_{0°(90°)}$ represents the normalized transmission under x(y)-polarized light. This means that significant transmission of x-polarized light is allowed [Fig. 5(b)]. In addition, rotating the magnetic field to $\varphi = 45°$ rotates the polarization axes accordingly. The emission peaks shift to $\pm 45°$ polarizations, and $LD_{45°} \approx 0.91$ at 347.85 THz [Fig. 5(c)]. Similarly, $LD_{45°}$ is defined as $LD_{45°} = \frac{T_{45°} - T_{-45°}}{T_{45°} + T_{-45°}}$, where $T_{45°(-45°)}$ corresponds to the transmission of $45°(-45°)$-polarized light. This implies that $45°$-polarized light is almost entirely transmitted while the orthogonal component is reflected. These results confirm the MO PhC's capability for high-performance polarization-selective emission and transmission, suitable for nanolasers and nonreciprocal photonic devices.

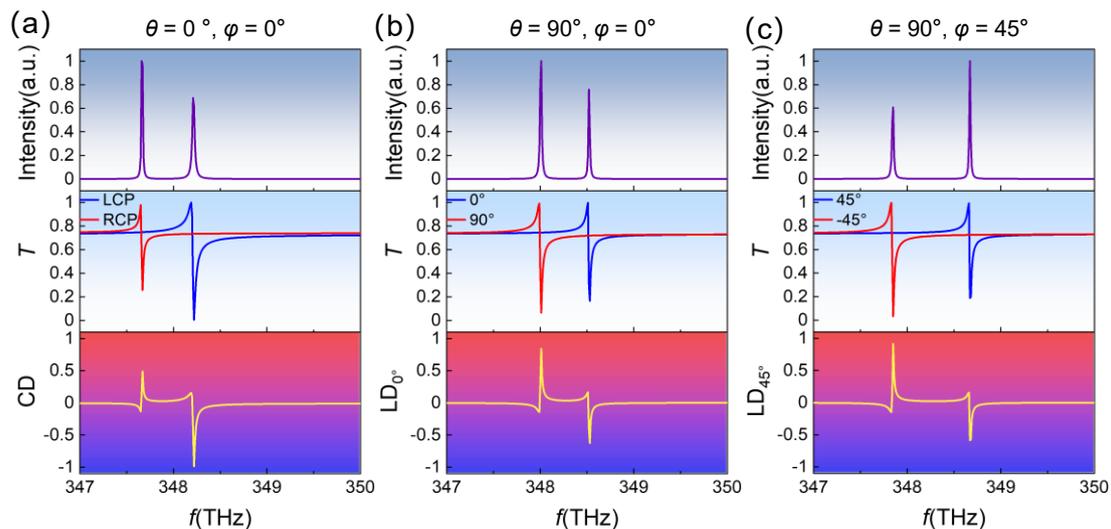

**FIG. 5.** Emission spectra (top), transmission spectra (middle), and dichroism (bottom) under different magnetic field directions: (a) out-of-plane magnetic field along the $+z$ direction ($\theta = 0°$), (b) in-plane magnetic field along the x direction($\theta = 90°$, $\varphi = 0°$), and (c) in-plane magnetic field along the $45°$ direction($\theta = 90°$, $\varphi = 45°$).

In the $C_6$ structure, the dipoles in the MO PhC can likewise achieve arbitrary far-field polarization by controlling the direction of the magnetic field, as shown in the Supplemental Material S7[38]. We also present simulations of the far-field polarizations of $|e_{1,2}^{C_6}\rangle$ in the $C_6$ structure over a certain k-space region under different magnetic field directions, as shown in the Supplemental Material S7[38]. This demonstrates the universality of our design: any structure with degenerate dipole modes can realize the same functionality.

**Conclusion:**

We introduce a new type of skyrmion, the MO skyrmion, which maps the real-space magnetic field orientation onto the far-field polarization of modes in the MO PhC. This mechanism constitutes a novel approach to realizing arbitrary polarization control of light. Magnetic field directions restricted to the $+x$ hemisphere are sufficient to realize all polarization states on the entire Poincaré sphere, yielding a MO polarization skyrmion with a skyrmion number of 2. The simulation results show excellent agreement with theoretical predictions. Beyond the $\Gamma$ point, polarizations at other k-space positions are also affected, with the influenced region expanding as the field strength increases. We further examine the evolution of the complex frequency and $Q$ factor with the magnetic field and find that BICs emerge under certain conditions when in-plane and out-of-plane fields are not applied simultaneously. Furthermore, the polarization textures of these BICs are governed by the magnetic field orientation, exhibiting features distinct from conventional symmetry-protected BICs. Strong polarization selectivity, with near-unity CD (~1) and high LD (>0.8), highlights promising applications. Our study establishes a versatile research platform that may open new possibilities in high-capacity optical communications, optical computing, imaging, and sensing.

**References:**

[1] C. He, H. He, J. Chang, B. Chen, H. Ma, and M. J. Booth, Light Sci Appl **10**, 194 (2021).

[2] Z. Zhan, M. Cantono, V. Kamalov, A. Mecozzi, R. Müller, S. Yin, and J. C. Castellanos, Science **371**, 931 (2021).

[3] Y. Deng, M. Wang, Y. Zhuang, S. Liu, W. Huang, and Q. Zhao, Light Sci Appl **10**, 76 (2021).

[4] Z. Zhu, D. Zhang, F. Xie, J. Ma, J. Chen, S. Gong, W. Wu, W. Cai, X. Zhang, M. Ren, and J. Xu, Optica **9**, 1297 (2022).

[5] L. Tong, X. Huang, P. Wang, L. Ye, M. Peng, L. An, Q. Sun, Y. Zhang, G. Yang, Z. Li, F. Zhong, F. Wang, Y. Wang, M. Motlag, W. Wu, G. J. Cheng, and W. Hu, Nat

Commun **11**, 2308 (2020).

[6] J. E. Solomon, Appl. Opt. **20**, 1537 (1981).

[7] Z. Shen, F. Zhao, C. Jin, S. Wang, L. Cao, and Y. Yang, Nat Commun **14**, 1035 (2023).

[8] M. Zhang, Q. Guo, Z. Li, Y. Zhou, S. Zhao, Z. Tong, Y. Wang, G. Li, S. Jin, M. Zhu, T. Zhuang, and S.-H. Yu, Science Advances **9**, eadi9944.

[9] W. Deng, L. Chen, H. Zhang, S. Wang, Z. Lu, S. Liu, Z. Yang, Z. Wang, S. Yuan, Y. Wang, R. Wang, Y. Yu, X. Wu, X. Yu, and X. Zhang, Laser & Photonics Reviews **16**, 2200136 (2022).

[10] P. Sen, J. V. Siles, N. Thawdar, and J. M. Jornet, Nature Electronics **6**, 164 (2023).

[11] J. C. Boileau, D. Gottesman, R. Laflamme, D. Poulin, and R. W. Spekkens, Phys. Rev. Lett. **92**, 017901 (2004).

[12] J. C. Boileau, R. Laflamme, M. Laforest, and C. R. Myers, Phys. Rev. Lett. **93**, 220501 (2004).

[13] C. Antonelli, M. Shtaif, and M. Brodsky, Phys. Rev. Lett. **106**, 080404 (2011).

[14] R. Duggan, J. del Pino, E. Verhagen, and A. Alù, Phys. Rev. Lett. **123**, 023602 (2019).

[15] H. Yang, H. Jussila, A. Autere, H.-P. Komsa, G. Ye, X. Chen, T. Hasan, and Z. Sun, ACS Photonics **4**, 3023 (2017).

[16] P. Weis, O. Paul, C. Imhof, R. Beigang, and M. Rahm, Appl. Phys. Lett. **95**, 171104 (2009).

[17] K. Y. Bliokh, C. T. Samlan, C. Prajapati, G. Puentes, N. K. Viswanathan, and F. Nori, Optica **3**, 1039 (2016).

[18] M. Born and W. Emil, *Principles of Optics* (Cambridge University Press, Cambridge University, 1999).

[19] R. Zhu, C. Qian, S. Xiao, J. Yang, S. Yan, H. Liu, D. Dai, H. Li, L. Yang, X. Chen, Y. Yuan, D. Dai, Z. Zuo, H. Ni, Z. Niu, C. Wang, K. Jin, Q. Gong, and X. Xu, Light Sci Appl **14**, 114 (2025).

[20] P. Oliwa, P. Kapuściński, M. Popławska, M. Muszyński, M. Król, P. Morawiak, R. Mazur, W. Piecek, P. Kula, W. Bardyszewski, B. Piętka, H. Sigurðsson, and J. Szczytko,

Advanced Science **12**, 2500060 (2025).

[21] X. Zhang, Y. Liu, J. Han, Y. Kivshar, and Q. Song, Science **377**, 1215 (2022).

[22] Y. Zhou, G. Zhao, Y. Du, and X. Ao, Physical Review B **109**, 075134 (2024).

[23] I. Katsantonis, A. C. Tasolamprou, E. N. Economou, T. Koschny, and M. Kafesaki, ACS Photonics **12**, 71 (2025).

[24] E. Mogni, G. Pellegrini, J. Gil-Rostra, F. Yubero, G. Simone, S. Fossati, J. Dostálek, R. Martínez Vázquez, R. Osellame, M. Celebrano, M. Finazzi, and P. Biagioni, Adv. Optical Mater. **10**, 2200759 (2022).

[25] X. Yang, J. Lv, J. Zhang, T. Shen, T. Xing, F. Qi, S. Ma, X. Gao, W. Zhang, and Z. Tang, Angewandte Chemie International Edition **61**, e202201674 (2022).

[26] X. Zhan, F.-F. Xu, Z. Zhou, Y. Yan, J. Yao, and Y. S. Zhao, Adv. Mater. **33**, 2104418 (2021).

[27] R. Zhang, X.-C. Li, and Q. H. Liu, Physical Review B **112**, 014417 (2025).

[28] X. Zhao, J. Wang, W. Liu, Z. Che, X. Wang, C. T. Chan, L. Shi, and J. Zi, Phys. Rev. Lett. **133**, 036201 (2024).

[29] W. Lv, H. Qin, Z. Su, C. Zhang, J. Huang, Y. Shi, B. Li, P. Genevet, and Q. Song, Science Advances **10**, eads0157 (2024).

[30] Y. Shen, Q. Zhang, P. Shi, L. Du, X. Yuan, and A. V. Zayats, Nat. Photon. **18**, 15 (2024).

[31] C. Guo, M. Xiao, Y. Guo, L. Yuan, and S. Fan, Phys. Rev. Lett. **124**, 106103 (2020).

[32] V. Hakobyan and E. Brasselet, Phys. Rev. Lett. **134**, 083802 (2025).

[33] P. Ornelas, I. Nape, R. de Mello Koch, and A. Forbes, Nat. Photon. **18**, 258 (2024).

[34] Y. Shen, E. C. Martínez, and C. Rosales-Guzmán, ACS Photonics **9**, 296 (2022).

[35] A. McWilliam, C. M. Cisowski, Z. Ye, F. C. Speirits, J. B. Götte, S. M. Barnett, and S. Franke-Arnold, Laser & Photonics Reviews **17**, 2300155 (2023).

[36] M. Król, H. Sigurdsson, K. Rechcińska, P. Oliwa, K. Tyszka, W. Bardyszewski, A. Opala, M. Matuszewski, P. Morawiak, and R. Mazur, Optica **8**, 255 (2021).

[37] Y.-P. Ruan, J.-S. Tang, Z. Li, H. Wu, W. Zhou, L. Xiao, J. Chen, S.-J. Ge, W. Hu, H. Zhang, C.-W. Qiu, W. Liu, H. Jing, Y.-Q. Lu, and K. Xia, Nat. Photon. **19**, 109 (2025).

[38] See Supplemental Material for details of derivations, and numerical examples.

[39] B. Zhen, C. W. Hsu, L. Lu, A. D. Stone, and M. Soljačić, Phys. Rev. Lett. **113**, 257401 (2014).

[40] N. Li, J. Zhang, D. N. Neshev, and A. A. Sukhorukov, ACS Photonics **12**, 1441 (2025).

[41] S. Wang, Z.-L. Deng, Y. Wang, Q. Zhou, X. Wang, Y. Cao, B.-O. Guan, S. Xiao, and X. Li, Light Sci Appl **10**, 24 (2021).

[42] L. Zhuang, J. Liu, Z. Jiang, Z. Ge, H. Ren, B. Cheng, X. Wei, and G. Song, Opt. Express **33**, 19282 (2025).